\documentclass[manuscript]{aastex}

\usepackage{amsmath}
\usepackage{threeparttable}

\newcommand\pder[2]{\ensuremath{\frac{\partial#1}{\partial#2}}}

\shorttitle{Damping in slow waves}
\shortauthors{S. K. Prasad, D. Banerjee, and T. Van Doorsselaere}
 
\begin{document}
\title{Frequency-dependent damping in propagating slow magneto-acoustic waves}             
\author{S. Krishna Prasad and D. Banerjee} 
\affil{Indian Institute of Astrophysics, II Block, Koramangala, Bangalore 560 034, India.}                                  
\email{krishna@iiap.res.in}
\author{T. Van Doorsselaere} 
\affil{Centre for mathematical Plasma Astrophysics, Mathematics Department, KU Leuven, Celestijnenlaan 200B bus 2400, 3001 Leuven, Belgium}                                  
 
\begin{abstract}
Propagating slow magneto-acoustic waves are often observed in polar plumes and active region fan loops. The observed periodicities of these waves range from a few minutes to few tens of minutes and their amplitudes were found to decay rapidly as they travel along the supporting structure. Previously, thermal conduction, compressive viscosity, radiation, density stratification, and area divergence, were identified to be some of the causes for change in the slow wave amplitude. Our recent studies indicate that the observed damping in these waves is frequency dependent. We used imaging data from SDO/AIA, to study this dependence in detail and for the first time from observations we attempted to deduce a quantitative relation between damping length and frequency of these oscillations. We developed a new analysis method to obtain this relation. The observed frequency dependence does not seem to agree with the current linear wave theory and it was found that the waves observed in the polar regions show a different dependence from those observed in the on-disk loop structures despite the similarity in their properties. 
\end{abstract}

\section{Introduction}
Polar plume/interplume regions and extended fan loop structures in active regions are often found to host outward propagating slow magneto-acoustic waves. Besides their contribution to coronal heating and solar wind acceleration, they are important for their seismological applications \citep{2003A&A...404L...1K, 2009ApJ...697.1674M, 2009A&A...503L..25W, 2011ApJ...727L..32V}. The observed periods of the slow waves are of the order of few minutes to few tens of minutes. These waves cause periodic disturbances in intensity and Doppler shift and are mostly identified from the alternate slanted ridges in the time-distance maps in intensity \citep{1998ApJ...501L.217D, 2000A&A...355L..23D}. However, spectroscopic studies by some authors indicate periodic asymmetries in the line profiles, suggesting the presence of high-speed quasi-periodic upflows, which also produce similar signatures in time-distance maps \citep{2010ApJ...722.1013D, 2011ApJ...727L..37T, 2011ApJ...738...18T}. This led to an ambiguity in the interpretation of observed propagating features as slow waves. But, later studies found that flow-like signatures are dominantly observed close to the foot points \citep{2011ApJ...737L..43N, 2012ApJ...759..144T} and no obvious blueward asymmetries were observed in the line profiles higher in the loops \citep{2012SoPh..281...67K, 2012A&A...546A..93G}. Results from the recent 3D magneto-hydrodynamic (MHD) simulations by \citet{2012ApJ...754..111O} and \citet{2013ApJ...775L..23W}, who report the excitation of slow waves by impulsively generated periodic upflows at the base of the coronal loop, were in agreement to this. Also, the propagation speeds were found to be temperature dependent for both sunspot \citep{2012SoPh..279..427K} and non-sunspot related structures \citep{2013ApJ...778...26U}, in agreement with the slow mode behaviour. So, the propagating disturbances observed in the extended loop structures and polar regions can be interpreted as due to slow waves.

One of the important observational characteristics of these waves is that they tend to disappear after travelling some distance along the supporting (guiding loop) structure. Their amplitude rapidly decays as they propagate. Thermal conduction, compressive viscosity, optically thin radiation, area divergence, and gravitational stratification, were identified to be some of the physical mechanisms that can alter the slow wave amplitude. The gravitational stratification leads to an increase in the wave amplitude whereas the other mechanisms cause a decrease \citep[see the review by][and the references therein]{2009SSRv..149...65D}. Using forward modelling to match the observed damping, it was found that for a slow mode with shorter (5~min) periodicity, thermal conduction is the dominant damping mechanism and when combined with area divergence it can account for the observed damping even when the density stratification is present \citep{2004A&A...415..705D}. They also found that the contribution of the compressive viscosity and radiative dissipation to this damping was minimal. Another study on oscillations with longer periods ($\approx$ 12 minutes) travelling along sub-million-degree cool loops, suggested that area divergence has the dominant effect over thermal conduction \citep{2011ApJ...734...81M}. Recently, \citet{2012A&A...546A..50K} had shown that this damping is dependent on frequency. These authors constructed powermaps in three different period ranges from which they conclude that longer period waves travel larger distances along the supporting structure while the shorter period waves get damped more heavily. Such frequency-dependent damping was earlier reported by \citet{2002ESASP.508..465W} and \citet{2002ApJ...580L..85O} for standing slow waves observed in hot coronal loops. In the present work, we aim to study the quantitative dependence of damping length of the wave on its frequency. Details on the observations are presented in the next section followed by the analysis method employed and the results obtained. Related theory and the physical implications of the results obtained are discussed in the subsequent sections.

\section{Observations}
Data used in this study are comprised of images taken by Atmospheric Imaging Assembly \citep[AIA;][]{2012SoPh..275...17L} on-board Solar Dynamics Observatory \citep[SDO;][]{2012SoPh..275....3P} in two different Extreme Ultra-Violet (EUV) channels centred at 171~\AA\ and 193~\AA. Full-disk images of three hours duration, starting from 21:10 UT on 2011 October 8, were considered. The cadence of the data is 12~s. The initial data at level 1.0, were processed to correct the roll angles and the data from different channels were brought to a common centre and common plate scale following the standard procedure using \verb+aia_prep.pro+ routine (version 4.13). The final spatial extent of each pixel is $\approx$ 0.6\arcsec.
\begin{figure}
 \centering
 \includegraphics[angle=90,width=7cm]{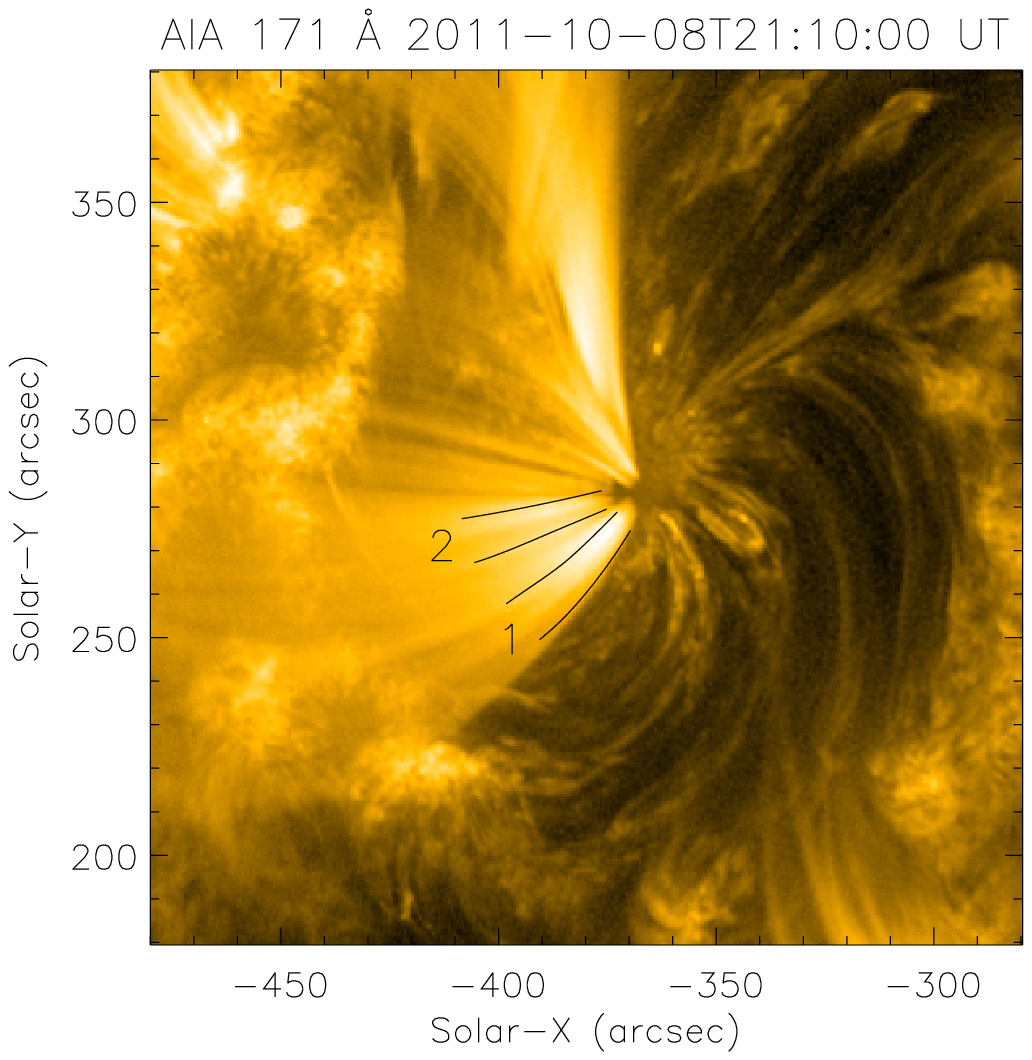}
 \includegraphics[angle=90,width=7.25cm]{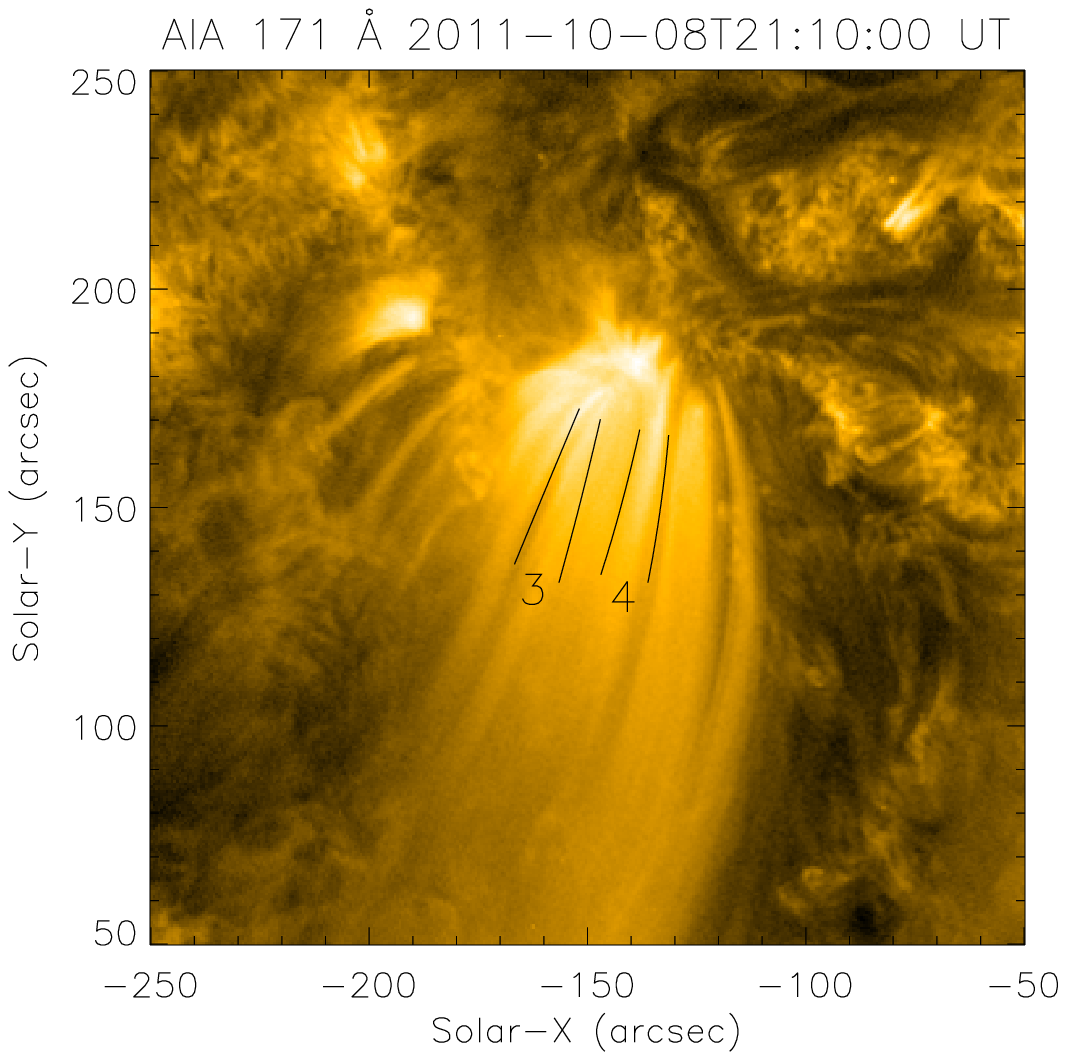}\\
 \hspace*{-0.43in}
 \includegraphics[angle=90,width=14.92cm]{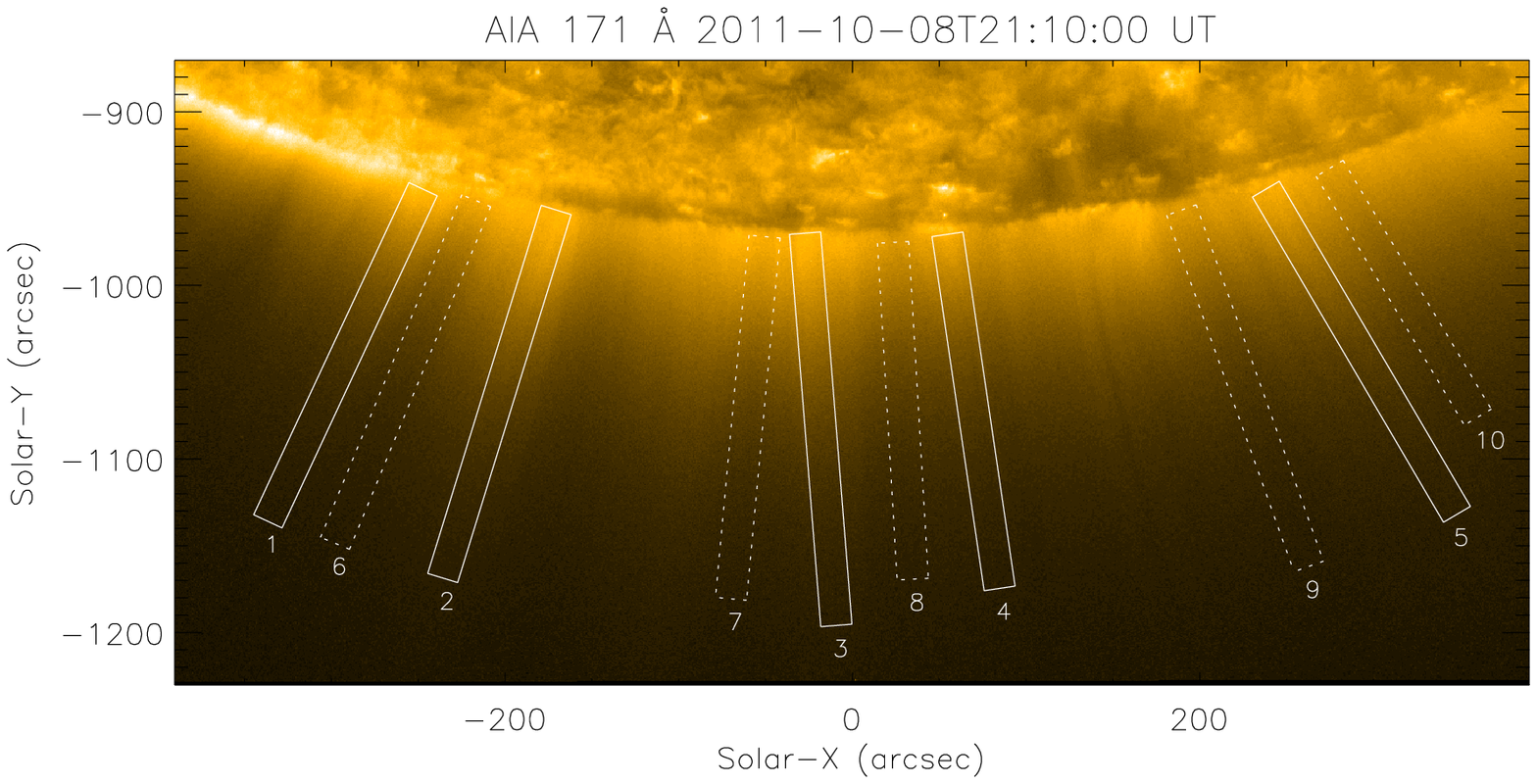} 
 \caption{Snapshots of the subfield regions chosen to cover a sunspot (left) and an on-disk plume like structure (right) and the plume/interplume structures at the south pole (bottom). Locations of the selected structures from these regions, are also marked.}
 \label{fig1}
\end{figure}

Subfield regions were chosen to cover loop structures over a sunspot, an on-disk plume like structure, and the plume/interplume structures at the south pole. The imaging sequence in each of these regions was co-aligned using intensity cross-correlation, taking the first image as the reference. A snapshot for each of the selected on-disk regions and the polar region are shown in Fig.~\ref{fig1}.

\section{Analysis \& Results}
Four loop structures, two from a sunspot region and another two from an on-disk plume like structure, were selected to represent the on-disk region and several plume and interplume regions at the south pole, were selected to represent the polar region, for this study. The selection of these structures was made on the basis of cleanliness of the propagating oscillations by looking at the time-distance maps. Fig.~\ref{fig1} displays the selected loop structures on-disk and the plume/interplume structures at the south pole. The width of the selected loop structures varied from 7 to 19 pixels and that of the plume/interplume structures was fixed at 30 pixels.
\begin{figure}
 \centering
 \includegraphics[angle=90,width=15.8cm]{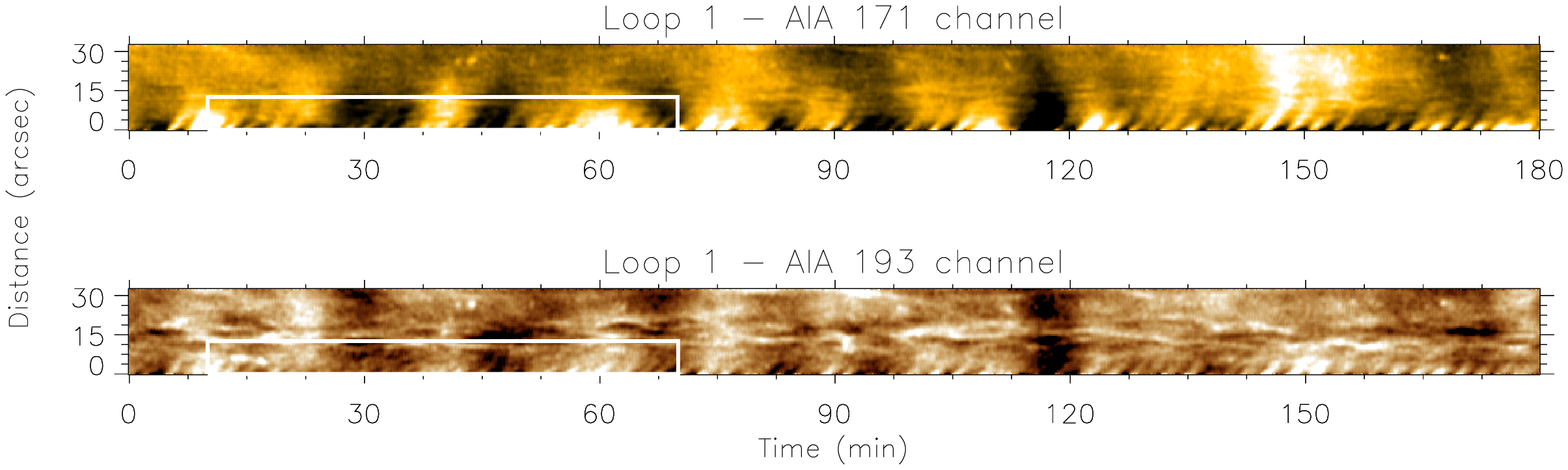}
 \includegraphics[angle=90,width=15.8cm]{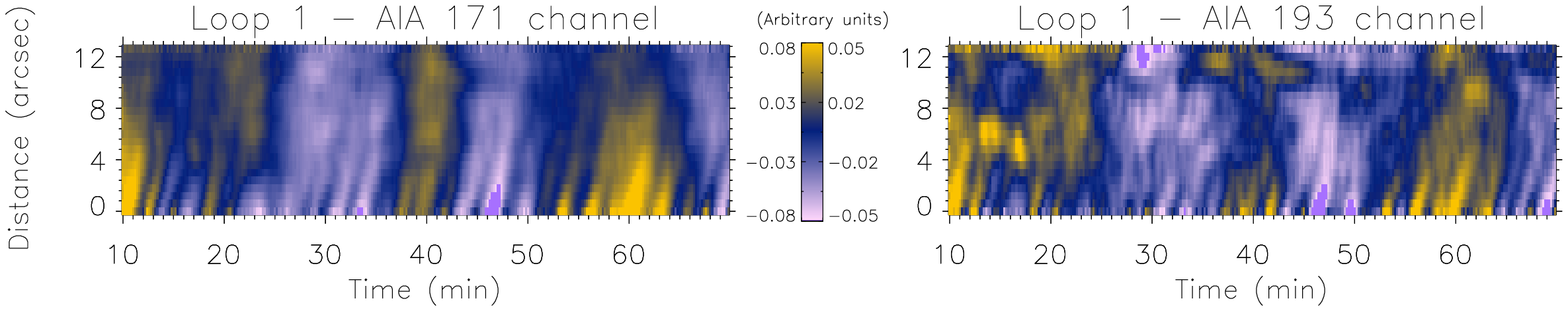} 
 \caption{Enhanced time-distance maps in 171 and 193 channels of AIA for the loop segment marked `1' in the top panel of Fig.~\ref{fig1}. The two panels in the bottom row display the region outlined with a box in the top two panels to present a zoomed-in view.}
 \label{fig2}
\end{figure}
The enhanced time-distance maps for the loop structure labelled `1' in the top panel of Fig.~\ref{fig1}, are shown in Fig.~\ref{fig2}, for both the AIA channels. A background constructed from the 300 point (60~min) running average in time has been subtracted from the original and the resultant is normalized with the same background to produce these enhanced time-distance maps. These maps clearly show alternate slanted ridges of varying intensity due to outward propagating slow waves. Ridges are not visible throughout the length of the chosen loop segment due to rapid decay in the slow wave amplitude as it propagates along the structure. However, they are present for the entire duration of the dataset. Another interesting feature visible in these maps is the presence of multiple periodicities. Two different periods, one less frequent (longer period) and another more frequent (shorter period), are apparent from these maps which can be more clearly seen from the zoomed-in view presented in the bottom panels of Fig.~\ref{fig2}. These shorter and longer periods roughly correspond to a periodicity of 3 and 22 minutes. There can be additional periods present in the signal which may not be not visually evident from these maps.

Our main aim is to measure the damping lengths of these waves at different periods and study the relation between them. Ideally one would look for measuring the damping lengths directly from the decaying wave amplitude along the loop, at a particular instant. But the damping in these waves is so rapid that we hardly get to observe more than a cycle. This makes the direct measurements difficult. The simultaneous presence of multiple periods is another hurdle for doing this. To overcome these issues, we transformed the original time-distance maps into period-distance maps by replacing the time series at each spatial position with its power spectrum. These maps contain the oscillation power at different periods for each spatial position. In this way, we can not only isolate the power in different periods, but can also trace the spatial decay in amplitude from the corresponding variation in power.
\begin{figure}
 \centering
 \includegraphics[angle=90,scale=0.42]{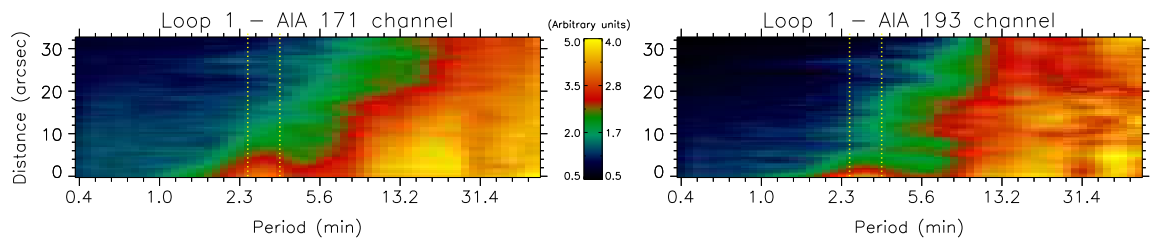}
 \caption{Period-distance maps in 171~\AA\ and 193~\AA\ channels of AIA generated from the time-distance maps for loop 1. Dotted lines enclose the power at 3~min period.}
 \label{fig3}
\end{figure}
Fig.~\ref{fig3} displays the period-distance maps generated from the time-distance maps for loop 1. A notable feature in these maps is the presence of more power in longer periods up to larger distances as observed by \citet{2012A&A...546A..50K}. 

\begin{figure}
 \includegraphics[angle=90,scale=0.5]{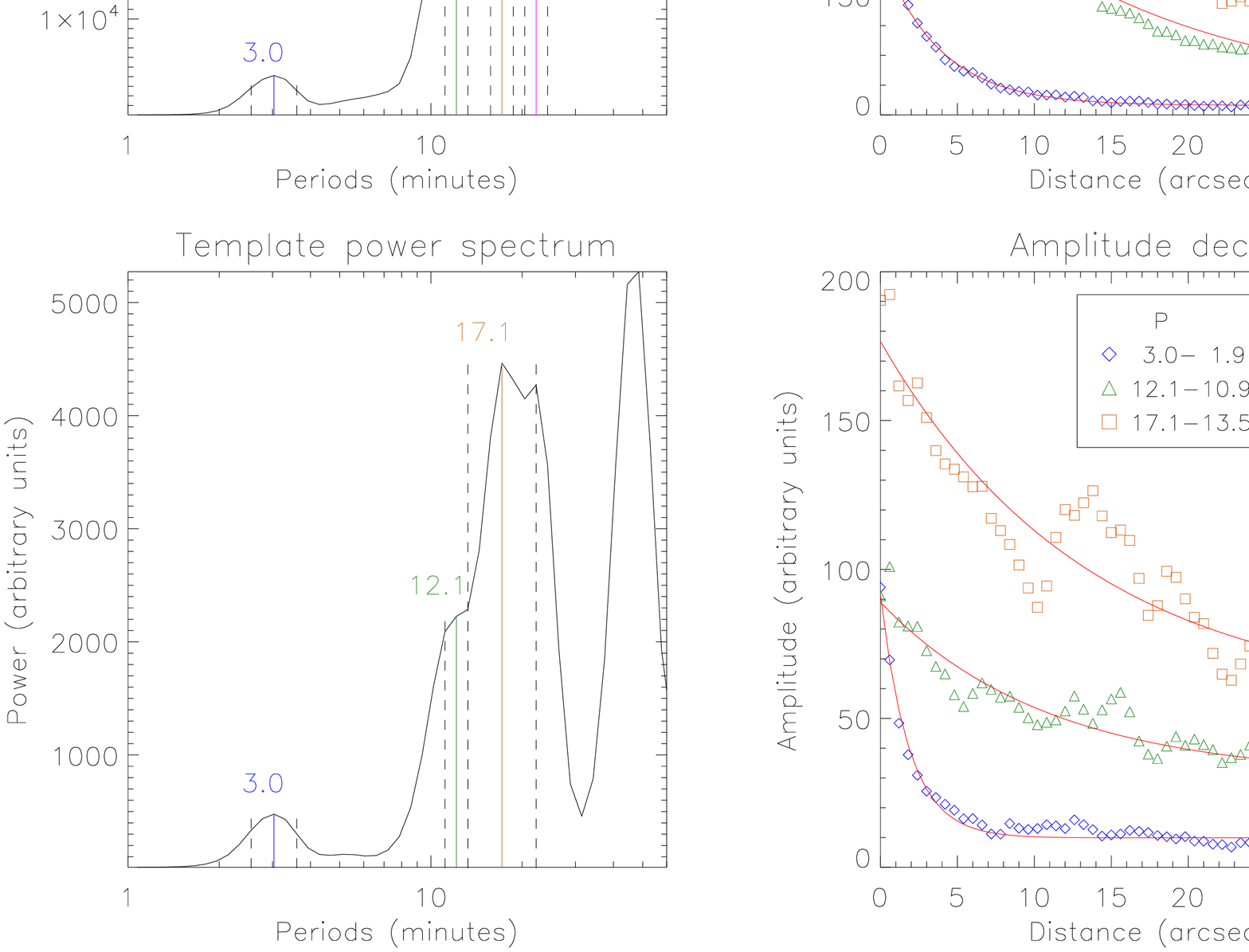}
 \caption{{\it Left:} Template power spectra constructed for loop 1. The vertical lines in solid, mark the identified peaks and the dashed lines mark their respective widths. Periodicities of individual peaks are listed above them in minutes. {\it Right:} Amplitude decay and the fitted model (solid line) for all the identified periods. Different symbols/colors were used to distinguish the different periods. A constant offset was added to the amplitudes to avoid cluttering between different periods. Damping lengths (in arcseconds) obtained from the fit and the corresponding reduced $\chi^{2}$ values are listed for each period in the plot legend. Top and bottom panels correspond to the data from 171 and 193 channels respectively.}
 \label{fig4}
\end{figure}
Now to identify all the periods present, we constructed an average light curve from the bottom 5 pixels of the structure and used it to generate a template power spectrum (see Fig.~\ref{fig4}). The peak periods and their respective widths were then estimated using a simple routine (\verb+gt_peaks.pro+) available with the solar software. At each peak identified, we constructed a bin of width determined by the width of the peak and computed the spatial variation of the total power in that bin from period-distance maps. Taking the square root of the power as amplitude of the oscillation, the amplitude decay at a particular period is fitted with a function of the form $A(y)=A_{0} e^{-y/L_{d}}+C$, to compute the damping length $L_{d}$ at that period. The template power spectrum constructed for loop 1 in the 171 channel, is shown in the left panel of Fig.~\ref{fig4}. All the identified peaks and their respective widths are marked with solid and dashed lines in this plot. Note the routine we used to estimate the widths of peaks gives very rough estimates which can be significantly different from the actual widths. But this is good enough to isolate the power in individual periods and is far better than the regular way of summing the power in predefined period ranges without having the knowledge of the peak frequencies present in the data. The amplitude decay and the fitted function corresponding to all the identified periods, are shown in the right panel of the figure. Different symbols (colors) are used to show the data for different periods. Corresponding plots for 193 channel are shown in the bottom panel. In the plots depicting the amplitude decay, the data for each period are offset by a constant value (50 for 171 and 5 for 193 channels) from the preceding period, to avoid cluttering. The computed damping lengths from each period are listed in the plot legend along with the respective errors obtained from the fit. The exact fit parameters ($A_{0}$, $L_{d}$, and $C$) estimated for all the periods obtained from the data are listed in Tables~\ref{par171} \& \ref{par193} of Appendix (Section~\ref{appendix}). The exponential fits are quite good for the amplitude decay in most of the cases, but occasionally we find some random variations (bumps) in the amplitude leading to very high damping lengths. We found that the contamination from background structures is causing this. To eliminate such data, we considered the damping lengths larger than the length of the supporting structure as unreliable. Thus, we measured the damping lengths at different periods. Only those periods between 2~min and 30~min were considered, keeping the total duration (3 hours) and cadence (12~s) of the dataset in mind. 
\begin{figure}
 \centering
 \includegraphics[angle=90,scale=0.65]{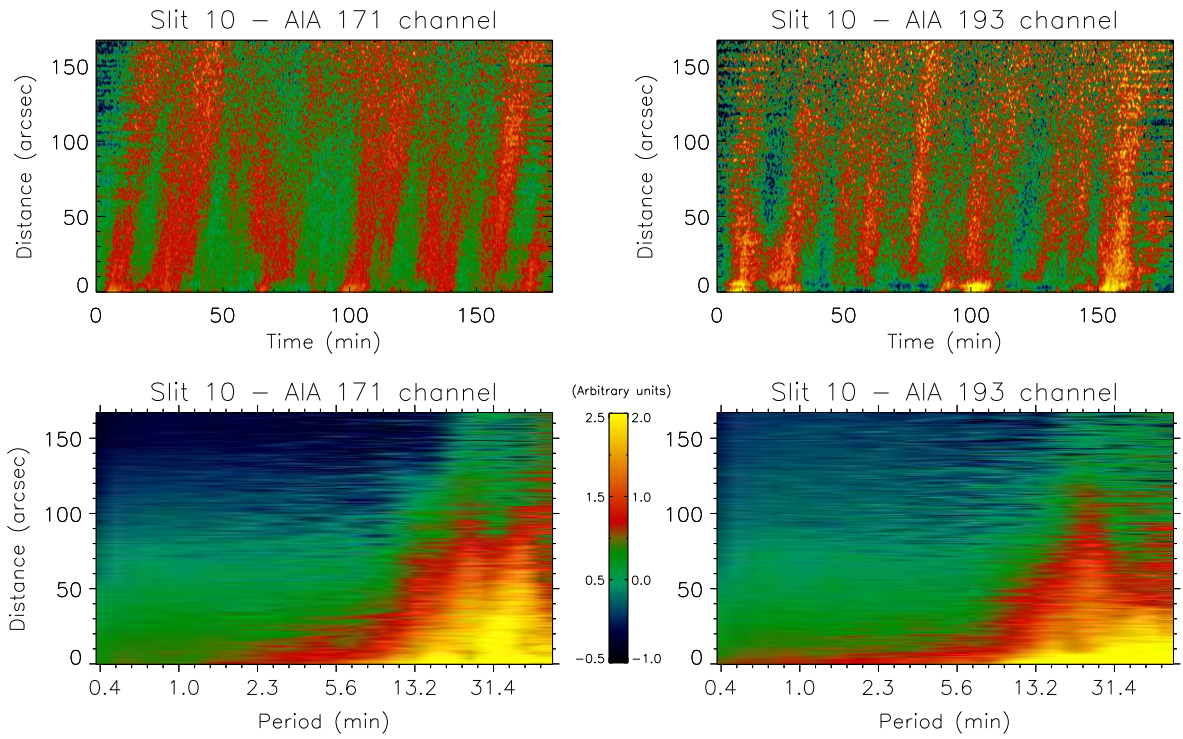}
 \caption{Enhanced time-distance maps (top) and period-distance maps (bottom) in 171 and 193 channels of AIA, generated from the interplume region marked by slit 10 in Fig.~\ref{fig1}.}
 \label{fig5}
\end{figure}

We combined the results from all the four loop structures on-disk and plotted the measured damping lengths against the period. This allows to evenly populate the frequency spectrum since the loops with different physical conditions support different frequencies. A similar procedure had been followed for the plume/interplume regions at the south pole except that the time-distance maps in the polar region are constructed by making artificial slits of 30 pixels (fixed) width to avoid the effect of jets \citep{2011A&A...528L...4K}. The time-distance maps and the corresponding period-distance maps constructed from the interplume region denoted by slit 10 (see Fig.~\ref{fig1}) are shown in Fig.~\ref{fig5} for both the channels. The propagating intensity disturbances are clearly seen in these images, but for some of the slit locations in 193 (slits 1, 2, 6, 7, \& 8, in Fig.~\ref{fig1}), the signal is very poor and we do not see any clear signature of these disturbances. This is possible because the 193 channel looks at relatively hotter plasma (1.25~MK) compared to the typical temperatures of plume/interplume regions ($<$1~MK). So the data from these locations are discarded in our final analysis. None of the data are discarded from 171 channel.
\begin{figure}
 \centering
 \includegraphics[angle=90,height=7cm]{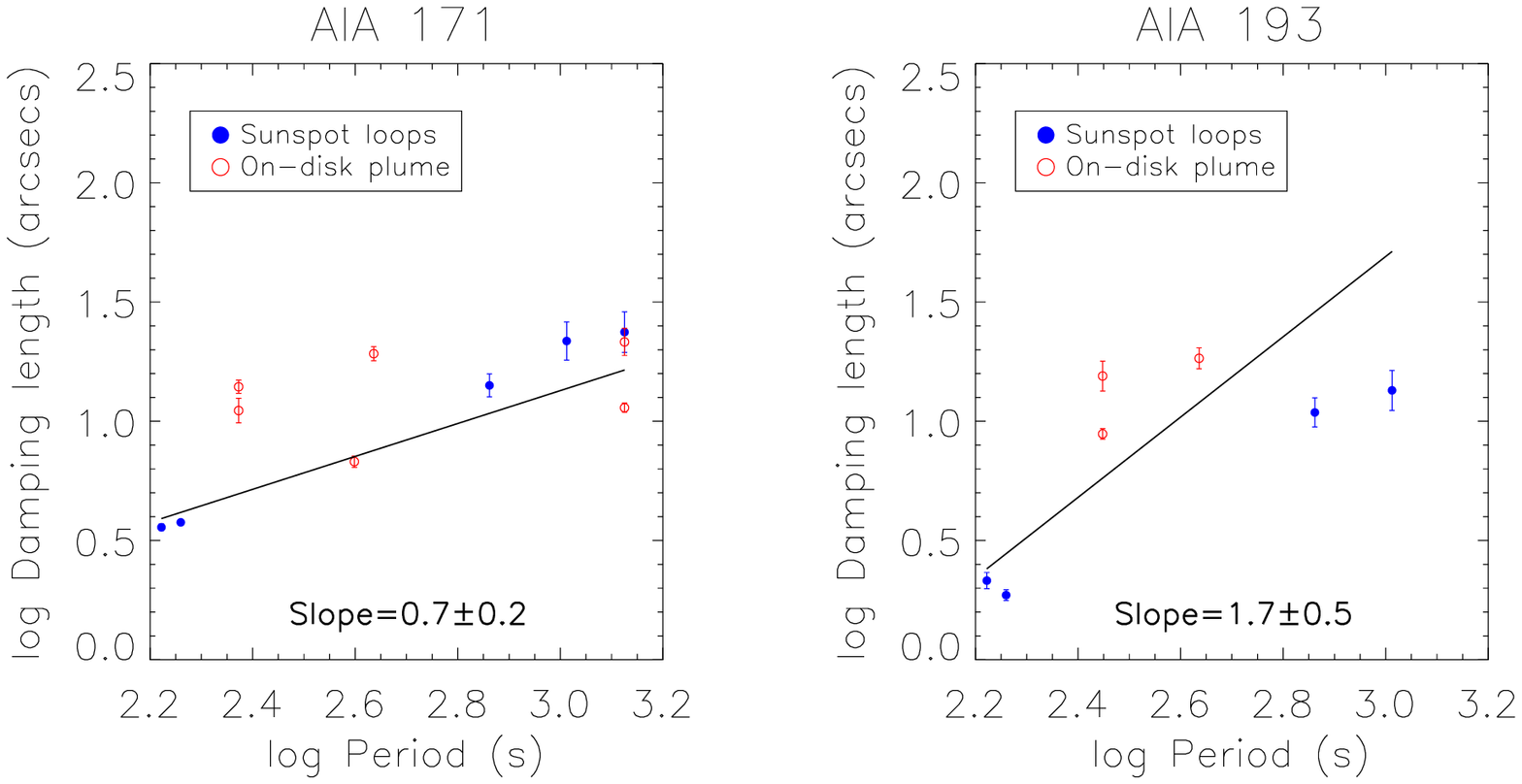}\\
\vspace*{0.2in}
 \includegraphics[angle=90,height=7cm]{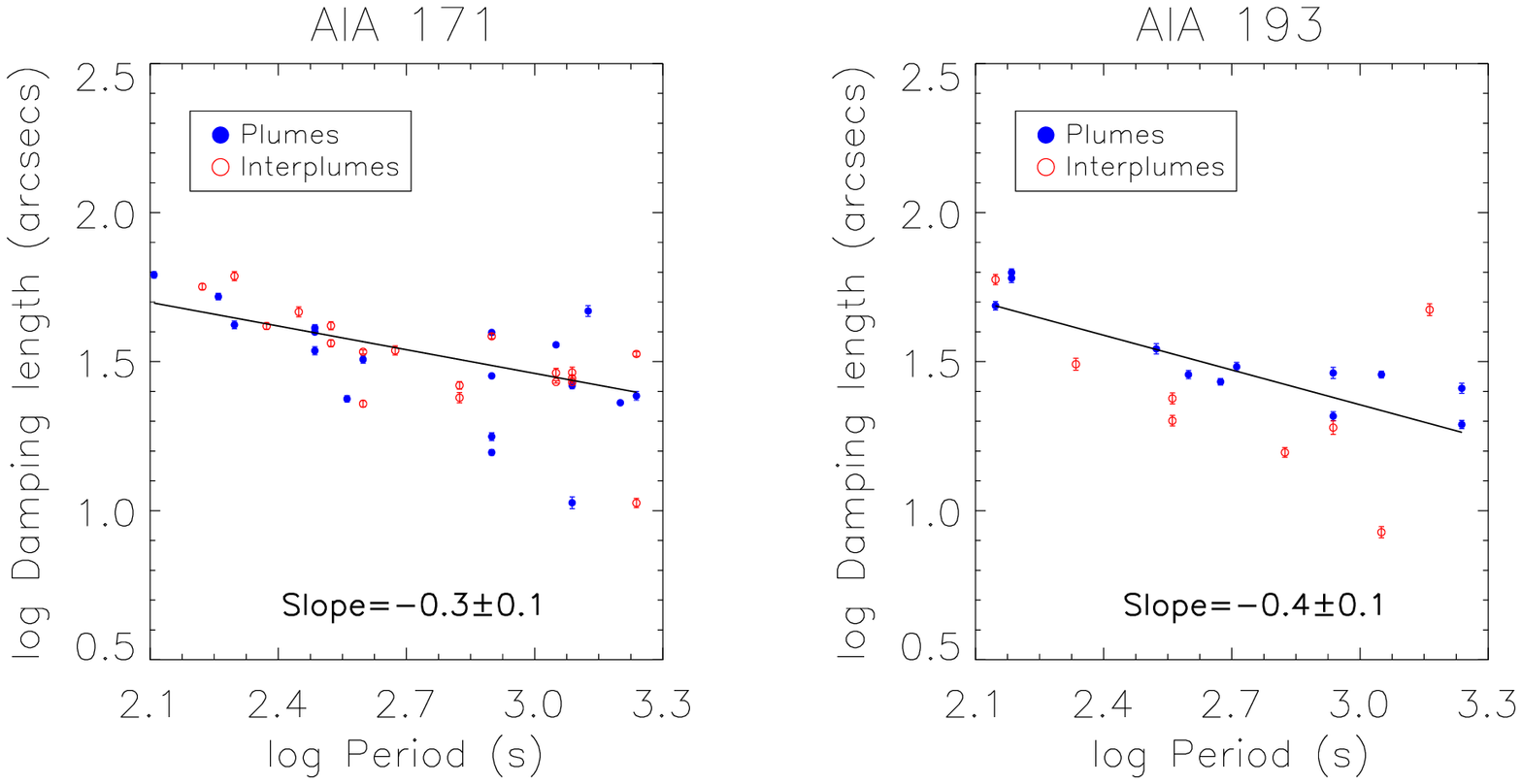}
 \caption{Frequency dependence of damping length for slow waves observed in loop structures on-disk (top) and in polar plume/interplume regions (bottom). Overplotted straight lines represent the linear fit. The slope of the line and the uncertainty in its estimate are written in the respective panels. Different symbols/colors correspond to different data as denoted in the respective legends.}
 \label{fig6}
\end{figure}
Fig.~\ref{fig6} displays the plots for damping length versus period in log-log scale. The top two panels correspond to the results from on-disk structures and the bottom two from polar regions. Different symbols (colors) are used to separate the data from sunspot \& plume-like structure and plume \& interplume regions. Damping lengths are measured in arcseconds and periods are measured in seconds. In all the panels, the overplotted solid lines represent a linear fit to the data. The slope of the line and the uncertainty in estimating it are written in the respective panels. The number of data points for the on-disk region are less because of the limited data. Clearly, the on-disk and polar regions show a different dependence of damping length on frequency.

\section{Theory}
\label{theory}
In this section, we study the theoretical dependence of the damping length on the frequency of the slow wave by considering different damping mechanisms separately. To perform this, we follow the one-dimensional linear MHD model of \citet{2003A&A...408..755D, 2004A&A...415..705D} and extend it to discuss the frequency dependence. This model is applicable under the assumptions that the magnetic field lines are straight, plasma-$\beta$ is much less than unity, and the amplitude of the oscillations are small. The one-dimensional form of the basic MHD equations for the slow waves can be written as
\begin{align}
& \pder{\rho}{t} = - \pder{}{z}(\rho v) \label{eq:cont}\\
& \rho \pder{v}{t} + \rho v \pder{v}{z} = - \pder{p}{z} - \rho g + \frac{4}{3} \eta_{0} \pder{^2v}{z^2} \label{eq:force}\\
& \pder{p}{t}+v \pder{p}{z} = - \gamma p \pder{v}{z} +(\gamma-1)\pder{}{z}\left(k_{\|}\pder{T}{z}\right) - (\gamma-1)[\rho^2\chi T^{\alpha} - H_{0}] \label{eq:energy}\\
& p = \frac{1}{\tilde{\mu}}\rho R T \label{eq:state}
\end{align}
where $p, \rho, v,$ and $T$ are pressure, density, velocity, and temperature respectively. $R$ is the gas constant and $\tilde{\mu}$ is the mean molecular weight. The second and third terms on the right hand side of Eq.~\ref{eq:force} represent the gravitational and viscous forces and those of Eq.~\ref{eq:energy} represent the energy losses due to thermal conduction and optically thin radiation. In these terms, $g$ is acceleration due to gravity, $\eta_{0}$ is coefficient of compressive viscosity, $k_{\|}$ is thermal conductivity parallel to the magnetic field, $\chi$ and $\alpha$ are constants under the approximation of a piece-wise continuous function for optically thin radiation, and $H_{0}$ is the coronal heating function. In the following subsections, we use appropriate forms of these equations to study the effect of individual damping mechanisms on slow waves and investigate the frequency dependence of damping length.

\subsection{Thermal conduction}
As the slow wave propagates energy is lost due to thermal conduction which results in a decay of its amplitude. By considering the thermal conduction as the only damping mechanism, we linearised the basic MHD equations and assumed the perturbations in the form exp$[i(\omega t - kz)]$, to obtain the following dispersion relation for the slow waves
\begin{equation}
 \omega^3-i\gamma d k^2 \omega^2 c_{s}^2 -\omega k^2 c_{s}^2 + i d k^4 c_{s}^4=0
 \label{eq:tcond}
\end{equation}
where $c_{s}$ is the adiabatic sound speed given by $c_{s}^2=\frac{\gamma p_{0}}{\rho_{0}}$ and $d$ is the thermal conduction parameter defined as $d=\frac{(\gamma-1)k_{\|}T_{0}}{\gamma c_{s}^2p_{0}}$. $p_{0}, \rho_{0},$ and $T_{0}$ are the equilibrium values of pressure, density, and temperature. The damping length of a propagating velocity perturbation of the form exp$[i(\omega t - kz)]$ is given by the reciprocal of the imaginary part of $k$. So, to solve for $k$, we simplify the dispersion relation by approximating the thermal conduction at its lower and upper limits. In the lower thermal conduction limit ($d\omega \ll$ 1) Eq.~\ref{eq:tcond} reduces to
\begin{equation}
k=\frac{\omega}{c_{s}} -i \frac{d\omega^2}{2 c_{s}} (\gamma-1)
\end{equation}
which gives the damping length $L_{d} \propto 1/\omega^2$. This implies in the lower thermal conduction limit, that the damping length of slow waves increases with the square of the wave period. Similarly, if we consider the higher thermal conduction limit ($d\omega \gg$ 1), the solution becomes
\begin{equation}
k=\gamma^{1/2}\frac{\omega}{c_{s}}-i \frac{\gamma-1}{2d\gamma^{3/2}c_{s}}
\end{equation} 
which gives the imaginary part of $k$ independent of $\omega$. Thus, in the limit of higher thermal conduction the damping in slow waves is frequency independent.
\subsection{Compressive viscosity}
The viscous forces lead to dissipation of energy and therefore reduce the slow wave amplitude. To understand the effect of compressive viscosity quantitatively, we solved the relevant linearised MHD equations assuming all the perturbations are in the form exp$[i(\omega t - kz)]$, which resulted in the following expression for wave number $k$
\begin{equation}
 k = \frac{\omega}{c_{s}}-i \frac{2}{3}\frac{\eta_{0} \omega^{2}}{\rho_{0} c_{s}^3}
\label{eq:visc}
\end{equation}
The imaginary part of this expression indicates that the amplitude of the slow wave decreases due to compressive viscosity and the decay lengths are proportional to the square of the wave period. A similar dependence was earlier reported by \citet{2000ApJ...533.1071O}.
\subsection{Optically thin radiation}
Energy dissipation due to radiation also causes a decay in wave amplitude. By retaining the radiation term and removing other dissipative terms in the basic MHD equations, one can obtain the dispersion relation for slow waves as 
\begin{equation}
 k=\frac{\omega}{c_{s}} -i\frac{r_{p}}{c_{s}},
\label{eq:rad}
\end{equation}
under the linear regime, for perturbations of the form exp$[i(\omega t - kz)]$. Here $r_{p}$ is the radiation parameter defined as $r_{p}=\frac{(\gamma-1)\rho_{0}^2\chi T_{0}^{\alpha}}{\gamma p_{0}}$. The reciprocal of this parameter has the dimension of time and gives the radiation time scale. According to Eq.~\ref{eq:rad}, the damping in slow waves due to optically thin radiation is frequency independent.
\subsection{Gravitational stratification}
\label{gravstrat}
In contrast to the other mechanisms so far discussed, the gravitational force stratifies the atmosphere which leads to an increase in the slow wave amplitude as it propagates outwards. Assuming the initial perturbations of the form exp$[i(\omega t - kz)]$, we solved the linearised MHD equations to get 
\begin{equation}
 k= i \frac{1}{2 H} \pm \frac{1}{c_{s}} \sqrt{\omega^2-\omega_{c}^2}
\end{equation}
Here $H$ is the gravitational scale height given by $H=\frac{p_{0}}{\rho_{0}g}$ and $\omega_{c}$ is the cut-off frequency defined as $\omega_{c}=\frac{g\gamma}{2c_{s}}$. This relation indicates that for slow waves with frequencies above the cut-off value $\omega_{c}$, the velocity amplitude grows exponentially as $e^{\frac{z}{2H}}$ and the growth rate is independent of frequency. The corresponding amplitude of density perturbations, however, varies as $e^{\frac{-z}{2H}}$ considering the equilibrium density fall $\propto e^{\frac{-z}{H}}$ due to stratification. Note that this variation still represents a growth in \textit{relative} amplitude as $e^{\frac{z}{2H}}$, similar to that of velocity perturbations and is independent of frequency.
\subsection{Magnetic field divergence}
\citet{2004A&A...415..705D} studied the effect of the radial divergence and area divergence of the magnetic field on slow waves. The amplitude of slow waves was found to decrease in both the cases as they propagate outwards. However, it is important to note that it is purely a geometric effect and there is no real dissipation mechanism involved. We solved the linearised MHD equations in the presence of radial divergence and obtained the following expression for the evolution of velocity perturbations
\begin{equation}
 v(r,t)=\mathrm{sin}(\omega t)\left[C_{1}j_{1}\left(\frac{r\omega}{c_{s}}\right)+C_{2}y_{1}\left(\frac{r\omega}{c_{s}}\right)\right],
\label{eq:magdiv}
\end{equation}
\citep[a similar expression was obtained by][]{2004A&A...415..705D}. Here $r$ is the radial coordinate in the spherical coordinate notation with the Sun at the centre, and $j_{1}(r\omega/c_{s})$ and $y_{1}(r\omega/c_{s})$ are first order spherical Bessel functions. Substituting the spherical Bessel functions with their standard definition, Eq.~\ref{eq:magdiv} can be written as
\begin{equation}
v(r,t)=\mathrm{sin}(\omega t)\left[C_1 \left(\frac{\mathrm{sin}\left(\frac{r\omega}{c_{s}}\right) - \frac{r\omega}{c_{s}}~\mathrm{cos}\left(\frac{r\omega}{c_{s}}\right)}{\left(\frac{r\omega}{c_{s}}\right)^2}\right) - C_2 \left(\frac{\mathrm{cos}\left(\frac{r\omega}{c_{s}}\right) + \frac{r\omega}{c_{s}}~\mathrm{sin}\left(\frac{r\omega}{c_{s}}\right)}{\left(\frac{r\omega}{c_{s}}\right)^2}\right)\right].
\end{equation}
The constants $C_{1}$ and $C_{2}$ can be determined from the boundary conditions. We chose these constants such that the amplitude of oscillations at the surface ($r=R_{\sun}$) is independent of frequency similar to that we assumed for other cases. Substituting $C_1=1/j_1(R_{\sun}\omega/c_{s})$ and $C_2=1/y_1(R_{\sun}\omega/c_{s})$, the velocity $v(r,t)$ becomes
\begin{equation}
v(r,t)= \mathrm{sin}(\omega t)\left[\frac{R_{\sun}^2}{r^2}\left(\frac{\mathrm{sin}\left(\frac{r\omega}{c_{s}}\right) - \frac{r\omega}{c_{s}}~\mathrm{cos}\left(\frac{r\omega}{c_{s}}\right)}{\mathrm{sin}\left(\frac{R_{\sun}\omega}{c_{s}}\right)-\frac{R_{\sun}\omega}{c_{s}}~\mathrm{cos}\left(\frac{R_{\sun}\omega}{c_{s}}\right)}\right) - \frac{R_{\sun}^2}{r^2}\left(\frac{\mathrm{cos}\left(\frac{r\omega}{c_{s}}\right) + \frac{r\omega}{c_{s}}~\mathrm{sin}\left(\frac{r\omega}{c_{s}}\right)}{\mathrm{cos}\left(\frac{R_{\sun}\omega}{c_{s}}\right)+\frac{R_{\sun}\omega}{c_{s}}~\mathrm{sin}\left(\frac{R_{\sun}\omega}{c_{s}}\right)}\right)\right]
\end{equation}
It can be shown that the amplitudes of expressions in the numerator varies as $r\omega/c_{s}$ and that in denominator varies as $R_{\sun}\omega/c_{s}$. This gives the overall amplitude variation as $1/r$ which is frequency independent. Following the same treatment, area divergence can be shown to behave similarly. Therefore, we can conclude that the damping in slow waves due to magnetic field divergence is frequency independent.
\begin{table}
\begin{center}
\caption{Dependence of damping length on period of slow waves}
\label{tab:theory}
\begin{tabular}{c c c}
\hline\hline
 Physical mechanism & Amplitude growth of  & Period dependence of \\
                    & density perturbations & damping length ($L_{d} \propto$) \\
 \hline
 Thermal conduction  &  & \\
  lower limit  & $e^{-\frac{d\omega^2}{2c_{s}}(\gamma-1)z}$  & $P^{2}$ \\
  upper limit  & $e^{-\frac{(\gamma-1)z}{2d\gamma^{3/2}c_{s}}}$  & $P^{0}$ \\
 Compressive viscosity & $e^{-\frac{2}{3}\frac{\eta_{0}\omega^2z}{\rho_{0}c_{s}^3}}$  & $P^{2}$ \\
 Optically thin radiation& $e^{-\frac{r_{p}z}{c_{s}}}$  & $P^{0}$ \\
 Gravitational stratification\tablenotemark{1} & $e^{\frac{-z}{2H}}$  & $P^{0}$ \\
 Magnetic field divergence & $R_{\sun}/r$  & $P^{0}$ \\
\hline
\end{tabular}
\tablenotetext{1}{Note the relative amplitude still grows. See Section~\ref{gravstrat} for details.}
\tablecomments{P is time period of the oscillation.}
\end{center}
\end{table}

A summary on the derived frequency dependence of damping length in slow waves is presented in Table~\ref{tab:theory} for different physical mechanisms. The table also lists the amplitude growth of density perturbations. It may be noted that although the derivations were primarily done for the velocity perturbations, the density (intensity) perturbations due to slow waves are proportional to the velocity perturbations (as can be derived from Eq.~\ref{eq:cont}), and hence the same growth is expected except for the case of gravitational stratification as mentioned in Section~\ref{gravstrat}. We did not explore the frequency dependence due to other geometrical effects like loop curvature, offset, and inclination, and other damping mechanisms like phase mixing, and resonant absorption, as we believe the damping in slow waves due to these effects is secondary. For instance, \citet{2004A&A...425..741D} studied the damping of slow waves due to phase mixing and mode coupling to the fast wave, using a two dimensional model and found that their contributions are not significant enough to explain the observed damping. 
\begin{figure}
 \centering
 \includegraphics[angle=90,scale=0.5]{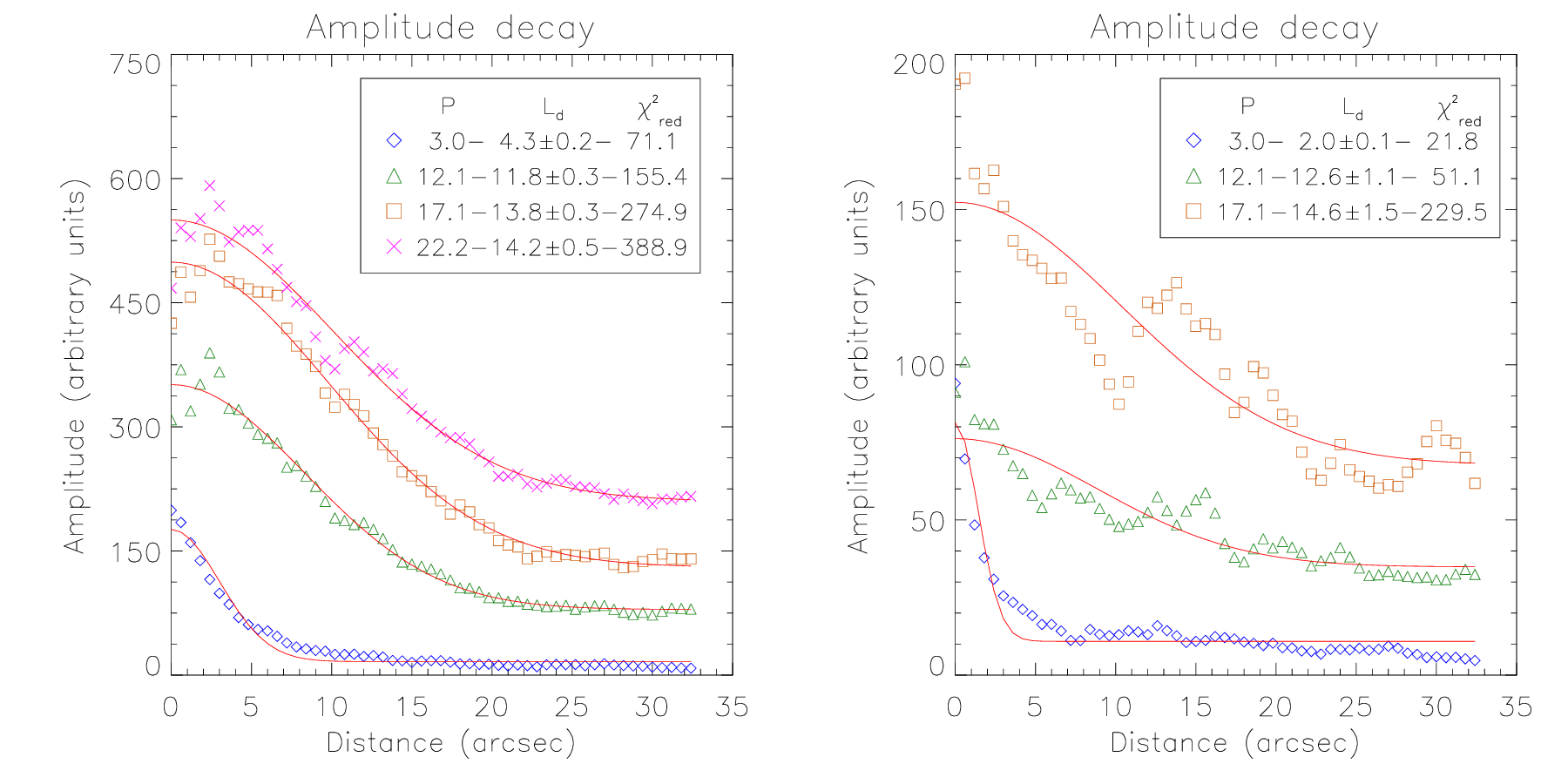}
 \caption{Amplitude decay for the periods shown in Fig.~\ref{fig4} fitted with a Gaussian decay model. The damping lengths (in arcseconds) and the corresponding reduced $\chi^{2}$ values are listed in the plot legend. The left and right panels show the results for 171 and 193 channels respectively.}
 \label{fig7}
\end{figure}
However, it may be interesting to note that the amplitude decay for some of the periods can be fit better with a Gaussian decay function ($A(y)=A_{0} e^{-y^2/L_{d}^2}+C$) rather than an exponential function (see Fig.~\ref{fig7}). A similar behaviour was found by \citet{2012A&A...539A..37P} in their numerical simulations for propagating kink waves. It was found analytically that the Gaussian damping for kink modes is a result of the excitation phase \citep{2013A&A...551A..39H}. In this phase other modes (than the kink mode) are excited and they gradually leak away before the system evolves to the ``eigenvalue'' state (when it oscillates with the pure kink mode). A consequence of this is that longer wavelengths show the Gaussian damping to greater heights. This fits also with some of our observations, where we find that the amplitude decay for the longer periods (and wavelengths) is quite well explained with Gaussian damping. It is unclear however, if the theory of \citet{2013A&A...551A..39H} for kink modes also holds for slow waves and what physical ingredients are essential for showing this behaviour.

\section{Discussion and Conclusions}
Damping in slow waves has been studied extensively in polar plumes and active region loops both theoretically and observationally since their first detection. However, studies on the frequency dependence of their damping are limited. \citet{2002ESASP.508..465W} and \citet{2002ApJ...580L..85O} studied the frequency-dependent damping in standing slow magneto-acoustic waves observed in hot ($T>$6~MK) coronal loops and found a good agreement between the observed scaling of the dissipation time with the period using their model. They concluded that thermal conduction is the dominant damping mechanism for these waves and the contribution of compressive viscosity is less significant. Theoretical investigations on frequency-dependent damping in propagating slow waves were made by a few authors \citep{2000ApJ...533.1071O, 2000A&A...362.1151N, 2001A&A...379.1106T}. Recently, \citet{2012A&A...546A..50K} report an observational evidence of this using powermaps constructed in three different period ranges. As a follow-up of that work, in this article we studied the quantitative dependence of damping lengths on frequency of the slow waves using period-distance maps.

We selected four loop structures on-disk and about 10 plume/interplume structures in the south polar region that show clear signatures of propagating slow waves. Damping lengths were measured and plotted against the period of the slow wave to find the relation between them. Fig.~\ref{fig6} displays the observed dependence of damping lengths on periodicity for the on-disk loop structures and the polar plume/interplume regions in two AIA channels. The slopes estimated from the linear fits are $0.7\pm0.2$ (171 channel) and $1.7\pm0.5$ (193 channel) for the on-disk regions and are $-0.3\pm0.1$ (171 channel) and $-0.4\pm0.1$ (193 channel) for the polar regions. The negative slopes obtained for the polar region means the damping lengths for the longer period waves observed in this region are shorter than those for the shorter period waves. Note, however, in both the regions the longer period waves are visible up to relatively larger distances due to the availability of more power. Considering thermal conduction, magnetic field divergence, and density stratification, as the dominant mechanisms that alter the slow wave amplitude, linear theory (see Table~\ref{tab:theory}) predicts the variation of damping length as square of the time period. In a log-log scale, used in Fig.~\ref{fig6}, this would mean a slope of 2. But as we find here, the slopes estimated from the observations are positive but less than 2 for the on-disk region and are negative for the polar region. It may be noted that similar negative slopes were found for the polar region, even when the data from plume and interplume regions were plotted separately. This mismatch between the observed values and those expected from the linear theory, suggests some missing element in the current theory of damping in slow waves. Perhaps, the linear description does not hold good and the slow waves undergo non-linear steepening that causes enhanced viscous dissipation \citep{2000ApJ...533.1071O}. This can be effective for the long period waves whose amplitudes are relatively larger and possibly can even explain the negative slopes observed in the polar regions. Further studies are required to explore such possibilities and understand the observed frequency dependence. Nevertheless, the discrepancy in the results from the on-disk and the polar regions, indicates the existence (or dominance) of different damping mechanisms in these two regions possibly due to different physical conditions. It is also possible that the sunspot loops and the on-disk plume-like structures also behave differently, but the current data is limited to make any such conclusions.

\acknowledgements 
We thank the anonymous referee for useful comments. The authors would also like to thank I. De Moortel for helpful discussions. The AIA data used here is the courtesy of SDO (NASA) and AIA consortium. This research has been made possible by the topping-up grant CHARM+top-up COR-SEIS of the BELSPO and the Indian DST. It was partly funded by the IAP P7/08 CHARM and an FWO Vlaanderen Odysseus grant.

\appendix
\section{Appendix}
\label{appendix}
The exponential fit parameters for all the periods identified in the on-disk and polar data are listed in the Tables~\ref{par171} \& \ref{par193} for 171 and 193 channels respectively. The corresponding reduced $\chi^{2}$ values are also listed as a goodness-of-fit statistic.

\begin{deluxetable}{c r@{.}l r@{$\pm$}l r@{$\pm$}l r@{$\pm$}l r@{.}l}
\tablewidth{0pt}
\tablecaption{Periodicity, exponential fit parameters, and the reduced $\chi^{2}$ values for all the data from AIA 171 channel \label{par171}}
\tablehead{\colhead{Structure} & \multicolumn{2}{c}{Period} & \multicolumn{2}{c}{$A_{0}$} & \multicolumn{2}{c}{$L_{d}$}  & \multicolumn{2}{c}{$C$} & \multicolumn{2}{c}{$\chi_{red}^2$} \\
  & \multicolumn{2}{c}{(min.)} & \multicolumn{2}{c}{(arbitrary units)} & \multicolumn{2}{c}{(arcsec)} & \multicolumn{2}{c}{(arbitrary units)} & \multicolumn{2}{c}{}}
\startdata
 Loop1    &     3&0   &   195.32&1.99    &    3.77&0.07     &   11.95&0.53     &    8&32     \\
 Loop1    &    12&1   &   373.84&13.66   &   14.14&1.57     &  -30.07&15.95    &  437&93    \\
 Loop1    &    17&1   &   573.04&46.96   &   21.71&4.01     & -128.19&54.29    & 1071&35    \\
 Loop1    &    22&2   &   549.66&51.02   &   23.65&4.63     & -107.21&58.02    &  885&18    \\
 Loop2    &     2&8   &   163.46&2.70    &   3.59 &0.11     &    9.04&0.69     &   14&83    \\
 Loop3    &     3&9   &    63.75&1.30    &   13.95&0.89     &    0.96&1.36     &    6&83    \\
 Loop3    &     7&2   &   155.38&3.76    &   19.21&1.31     &   -9.08&4.51     &   25&39    \\
 Loop3    &    22&2   &   275.87&4.35    &   11.39&0.52     &   19.04&3.54     &   85&67    \\
 Loop4    &     3&9   &    13.69&0.54    &   11.09&1.32     &    7.20&0.51     &    1&12    \\
 Loop4    &     6&6   &    32.25&0.80    &    6.77&0.37     &   10.30&0.37     &    2&09    \\
 Loop4    &    22&2   &   140.94&7.89    &   21.51&2.81     &   -3.37&9.20     &   37&76    \\
 Slit1    &     6&1   &    15.70&0.23    &   23.72&0.59     &    1.50&0.05     &    0&55    \\
 Slit1    &    13&2   &    39.83&0.30    &   28.33&0.38     &    1.60&0.08     &    1&11    \\
 Slit1    &    20&4   &    64.69&0.90    &   26.30&0.64     &    2.42&0.23     &    9&23    \\
 Slit2    &     2&1   &     7.59&0.07    &   61.83&1.52     &    0.86&0.05     &    0&11    \\
 Slit2    &     5&1   &    32.99&0.14    &   39.73&0.34     &    1.49&0.05     &    0&34    \\
 Slit2    &    13&2   &    57.21&0.31    &   39.62&0.43     &    0.92&0.12     &    1&61    \\
 Slit2    &    22&2   &    53.16&0.99    &   46.75&1.95     &    0.44&0.47     &   19&26    \\
 Slit3    &    13&2   &    60.78&0.79    &   15.68&0.32     &    2.55&0.13     &    4&45    \\
 Slit3    &    20&4   &    86.50&2.54    &   10.63&0.48     &    4.23&0.32     &   31&98    \\
 Slit3    &    28&8   &   103.80&2.08    &   24.27&0.82     &    4.89&0.46     &   46&10    \\
 Slit4    &     3&3   &     7.24&0.10    &   42.02&1.26     &    1.17&0.04     &    0&17    \\
 Slit4    &     5&1   &     9.55&0.15    &   34.41&1.08     &    1.28&0.05     &    0&35    \\
 Slit4    &     6&6   &    10.44&0.16    &   32.16&0.91     &    1.31&0.05     &    0&33    \\
 Slit4    &    13&2   &    24.97&0.47    &   17.73&0.54     &    2.02&0.09     &    1&76    \\
 Slit4    &    26&4   &    54.65&0.48    &   23.02&0.34     &    2.02&0.11     &    2&28    \\
 Slit5    &     3&0   &     7.02&0.07    &   52.24&1.33     &    1.20&0.04     &    0&11    \\
 Slit5    &     5&1   &    11.39&0.14    &   41.03&1.06     &    1.81&0.06     &    0&35    \\
 Slit5    &    18&7   &    41.99&0.24    &   36.01&0.40     &    1.51&0.08     &    0&88    \\
 Slit6    &     3&9   &     7.43&0.09    &   41.63&1.09     &    1.10&0.04     &    0&15    \\
 Slit6    &     6&6   &    13.09&0.20    &   22.84&0.58     &    1.48&0.04     &    0&39    \\
 Slit6    &    18&7   &    57.04&0.30    &   27.03&0.25     &    1.94&0.08     &    1&04    \\
 Slit6    &    28&8   &    40.15&0.39    &   33.55&0.62     &    2.50&0.13     &    2&17    \\
 Slit7    &     6&6   &    20.37&0.17    &   34.12&0.53     &    1.28&0.06     &    0&40    \\
 Slit7    &    20&4   &    62.27&1.44    &   29.08&1.21     &    4.00&0.41     &   26&00    \\
 Slit8    &     3&3   &     5.58&0.07    &   61.22&2.17     &    0.90&0.06     &    0&09    \\
 Slit8    &     5&6   &     7.75&0.10    &   41.73&1.28     &    1.20&0.05     &    0&19    \\
 Slit8    &     7&9   &     9.90&0.17    &   34.48&1.21     &    1.38&0.06     &    0&44    \\
 Slit8    &    13&2   &    18.98&0.17    &   38.47&0.73     &    1.12&0.07     &    0&46    \\
 Slit8    &    20&4   &    21.98&0.30    &   27.80&0.70     &    1.86&0.09     &    1&10    \\
 Slit8    &    28&8   &    35.96&0.84    &   10.61&0.39     &    3.60&0.12     &    3&51    \\
 Slit9    &     2&8   &     6.51&0.06    &   56.43&1.37     &    1.37&0.04     &    0&08    \\
 Slit9    &     5&6   &     8.24&0.11    &   36.46&0.97     &    1.30&0.04     &    0&20    \\
 Slit9    &    11&1   &    18.76&0.32    &   26.31&0.77     &    1.66&0.08     &    1&17    \\
 Slit9    &    20&4   &    50.68&0.30    &   26.99&0.28     &    1.63&0.08     &    1&07    \\
Slit10    &     4&7   &     7.03&0.10    &   46.45&1.78     &    1.33&0.07     &    0&17    \\
Slit10    &    11&1   &    17.83&0.39    &   23.94&0.96     &    2.44&0.11     &    1&60    \\
Slit10    &    18&7   &    45.65&0.81    &   28.99&1.03     &    3.92&0.30     &    8&27    \\
\enddata
\end{deluxetable}

\begin{deluxetable}{c r@{.}l r@{$\pm$}l r@{$\pm$}l r@{$\pm$}l r@{.}l}
\tablewidth{0pt}
\tablecaption{Periodicity, exponential fit parameters, and the reduced $\chi^{2}$ values for all the data from AIA 193 channel \label{par193}}
\tablehead{\colhead{Structure} & \multicolumn{2}{c}{Period} & \multicolumn{2}{c}{$A_{0}$} & \multicolumn{2}{c}{$L_{d}$}  & \multicolumn{2}{c}{$C$} & \multicolumn{2}{c}{$\chi_{red}^2$} \\
  & \multicolumn{2}{c}{(min.)} & \multicolumn{2}{c}{(arbitrary units)} & \multicolumn{2}{c}{(arcsec)} & \multicolumn{2}{c}{(arbitrary units)} & \multicolumn{2}{c}{}}
\startdata
 Loop1   &    3&0   &    81.62&2.39    &    1.87&0.10    &    9.90&0.43    &    7&84    \\
 Loop1   &   12&1   &    58.75&2.72    &   10.89&1.53    &   25.23&2.60    &   27&39    \\
 Loop1   &   17&1   &   121.21&7.62    &   13.48&2.61    &   45.37&8.72    &  154&00    \\
 Loop2   &    2&8   &    63.19&2.82    &    2.15&0.17    &    8.05&0.54    &   11&76    \\
 Loop3   &    4&7   &    22.28&0.47    &    8.85&0.45    &    5.72&0.26    &    0&89    \\
 Loop3   &    7&2   &    37.66&1.31    &   18.37&1.86    &    6.90&1.57    &    3&61    \\
 Loop4   &    4&7   &    13.32&0.65    &   15.48&2.23    &    4.98&0.77    &    0&88    \\
 Slit3   &    2&5   &     4.54&0.04    &   62.96&1.72    &    0.81&0.03    &    0&05    \\
 Slit3   &    7&9   &    13.74&0.20    &   27.08&0.68    &    1.30&0.05    &    0&48    \\
 Slit3   &   14&4   &    26.11&0.65    &   28.98&1.26    &    1.60&0.17    &    5&27    \\
 Slit3   &   28&8   &    70.20&1.39    &   19.46&0.62    &    3.47&0.26    &   16&59    \\
 Slit4   &    2&3   &     3.48&0.05    &   48.70&1.59    &    0.98&0.03    &    0&04    \\
 Slit4   &    5&6   &     5.01&0.10    &   34.91&1.38    &    1.11&0.04    &    0&15    \\
 Slit4   &    8&6   &     7.31&0.13    &   30.40&1.03    &    1.17&0.04    &    0&24    \\
 Slit4   &   14&4   &    14.76&0.31    &   20.76&0.72    &    1.48&0.07    &    0&87    \\
 Slit4   &   28&8   &    29.39&0.67    &   25.74&1.02    &    1.87&0.17    &    4&97    \\
 Slit5   &    2&5   &     2.55&0.03    &   60.29&2.05    &    0.93&0.02    &    0&02    \\
 Slit5   &    6&6   &     7.11&0.13    &   28.61&0.92    &    1.77&0.03    &    0&21    \\
 Slit5   &   18&7   &    17.68&0.26    &   28.62&0.74    &    1.84&0.07    &    0&82    \\
 Slit9   &    2&3   &     2.11&0.03    &   59.64&2.32    &    0.90&0.02    &    0&02    \\
 Slit9   &    6&1   &     4.13&0.10    &   23.80&1.00    &    1.11&0.02    &    0&11    \\
 Slit9   &   11&1   &     9.54&0.23    &   15.69&0.59    &    1.28&0.04    &    0&36    \\
 Slit9   &   14&4   &     8.02&0.26    &   19.00&1.01    &    1.37&0.05    &    0&58    \\
 Slit9   &   24&3   &     9.74&0.19    &   47.24&2.14    &    1.04&0.10    &    0&74    \\
Slit10   &    3&6   &     3.33&0.07    &   31.00&1.44    &    1.19&0.03    &    0&07    \\
Slit10   &    6&1   &     4.92&0.12    &   20.04&0.82    &    1.33&0.03    &    0&12    \\
Slit10   &   18&7   &    30.96&0.87    &    8.47&0.37    &    3.21&0.12    &    3&09    \\
\enddata                                                                             
\end{deluxetable}
                                                                                                                                                         

\end{document}